\documentstyle[12pt]{article}
\def\({\left (}
\def\){\right )}
\def\[{\left [}
\def\]{\right ]}
\def\be{\begin{equation}}
\def\ee{\end{equation}}
\def\hphi{\hat{\phi}}
\def\Tr{{\rm Tr}}
\begin{document}
 \begin{titlepage}
 \vspace*{-4ex}
 \null \hfill MPI-Ph/92-72\\
 \null \hfill hep-th/9209227 \\
 \null \hfill August, 1992 \\
\vskip 2.2 true cm
 \begin{center}
 {\bf \LARGE The 3d Effective Field Theory   \\[0.1cm]
         of the High Temperature Abelian Higgs Model
    }\\[11ex]
{\large
  Vidyut Jain }
   \\ [4ex]
%

Max-Planck-Institut f\"{u}r Physik \\
  Werner-Heisenberg-Institut  \\
 P.O.\,Box 40 12 12 , D - 8000 Munich 40, Germany \\ [2ex]
 \end{center}
\vskip 2.3cm
 \begin{abstract}
We study a weakly coupled
3 dimensional $\varphi^4$ type model consisting of $N$ real
scalars $\varphi^i$ coupled to an abelian gauge field ${\cal A}^a$ and
one extra scalar field $\rho$. We argue that, below some scale
$\Lambda=O(T)$,  this is the effective field
theory of a 3+1 dimensional abelian Higgs model at a high temperature T.
The effective theory is sufficient to study the nature of the
phase transition in four dimensions.
By introducing an auxiliary field $\chi$ we  eliminate the
explicit $\varphi^4$ term; the new Lagrangian allows for a simple
computation of the dominant corrections to the effective scalar
potential $V_{eff}$ in the large $N$ limit. We study three cases:
a) $6g^2/\lambda \leq O(1)$, b) $6g^2/\lambda\sim O(N)$ and
c) $6g^2/\lambda\sim O(N^{2\over3})$ where $g$ and $\lambda$ are the
4d gauge and scalar couplings, respectively. For case a) which  is
the most thoroughly studied we find that
the leading $O(N)+O(1)$ result for $V_{eff}$ admits only
a second order phase transition. For the other cases we find that
 b) the leading $O(1)$ result
for $V_{eff}$ admits a first order phase transition whose strength is
independent of $N$ and c) the leading $O(N^{1\over3})$ result for
$V_{eff}$ admits only a second order phase transition -- the $O(1)$
corrections to this can be interpreted as indicating a first order phase
transition whose strength diminishes as $N$ increases.
 \end{abstract}
 \end{titlepage}

\renewcommand{\thefootnote}{\alph{footnote}}

\noindent {\Large \bf 1. Introduction.}

There has been much recent interest in the nature of the electroweak
phase transition. In general, for a spontaneously broken 4d $\phi^4$
type model, one can attempt to compute the high temperature effective
scalar potential $V_{eff}$.
At very high temperature the symmetry is believed
to be restored, and
at some temperature $T=T_2$ the origin ($|\phi|=0$) goes
from being a local minimum to being a local maximum. If at $T>T_2$
the origin is also the global minimum the theory admits only a
second order phase transition which proceeds by a roll--over after
the temperature drops below $T_2$. On the other hand, if  at some
$T_1>T_2$ the scalar potential has another minimum, degenerate in
energy with the one at the origin, the theory can admit a first
order phase transition.

The determination of a reliable effective potential has proven
difficult because, in generic models, ordinary perturbation
theory in $\hbar$ has infrared divergence problems just at values
of $\phi^2$ and $T$ where the phase transition occurs. In a $\phi^4$
model with $N$ real scalars,
Dolan and Jackiw [1] showed over 18 years ago that the one--loop
potential has problems. They summed up an infinite class of
diagrams, the ``superdaisies'' to obtain a reliable result
for the effective
mass at the origin.

Giving a reliable estimate of $V_{eff}$
   in the case of the standard model is much more involved.
The typical approach [2,3,4,5] for the corrections from the
bosonic sector of the standard model has been to use not tree--level
propogators but propogators
corrected by one--loop vacuum polarization effects in a  one--loop
calculation of
the effective potential. This has the effect of summing up
an infinite class of diagrams, the so--called ``ring'' diagrams, with
the correct combinatorics except for the two--loop graphs [4,5,6]
which are overcounted.  When
only the leading $T^2$ dependent corrections are used for the vacuum
polarizations the overcounting is not serious, but when field
dependent vacuum polarizations are used it can result in dangerous
$T^3|\phi|$ terms in the effective potential [4,5] which results in
the effective mass blowing up as one approaches the origin.

Assuming no
such dangerous terms appear, previous studies have concentrated on
determining the coefficient of the $|\phi|^3$ term in
their estimates of $V_{eff}$. For small $\phi^2$, the effective
potential can be expanded as ($d>0$)
\be V_{eff} = c+ a\phi^2 - b|\phi|^3 + d\phi^4 +\cdots . \ee
The coefficient $a$ is positive when $T>T_2$ and negative when $T<T_2$.
In a general 3+1 dimensional
scalar and gauge system the coefficients $a$, $b$ and $d$
can be found in the
ring approximation to $V_{eff}$ as follows [2,3,4,5].
In the background
field evaluation of $V_{eff}$ one expands the action about
quantum scalar fields $\hphi^i$ and quantum gauge fields $A^\mu$.
Each quantum degree of freedom has a mass depending on the background
field $\phi^2$. At zero external momentum, the leading
vacuum polarization effects give additional $T^2$ contributions
to the mass of the quantum scalars and the longitudinal gauge
field, but not to the transverse gauge fields. Then, when $T$ is
sufficiently bigger than $T_2$,
the result of the  polarization effects is that the scalar and
longitudinal gauge boson loops  give  only $\phi^2$ and $\phi^4$
contributions to eq. (1). In this limit, the transverse gauge
fields still contribute a $|\phi|^3$ term. It is the existence
of this term in the context of the standard model which led
the authors of [2,4] to conclude that the electroweak model
with sufficiently small self--scalar coupling admits a (weak) first
order phase transition.

The analysis we have just described is however incomplete.
For example, how are these results modified if one keeps
the field dependent contributions to the vacuum polarizations?
As already mentioned, the simple ring diagram procedure can lead to
spurious $O(|\phi|)$ contributions to $V_{eff}$, so obviously
some care is needed. One would also like to investigate
how the momentum dependence of the vacuum polarizations modifies
the $|\phi^3|$ result. Finally, we would like to know in what way
the ring (or ``daisy'') diagram sum can be interpreted as  a
consistent approximation scheme.

To address these questions  and problems,
we study as a toy model a
3 dimensional $\varphi^4$ type model consisting of $N$ real
scalars $\varphi^i$ coupled to an abelian gauge field ${\cal A}^a$ and
an extra scalar $\rho$, which we argue is the low energy effective
field theory for a   3+1 dimensional Abelian Higgs model at
high temperature. The computations in the effective 3d theory
are however much easier than in the full four dimensional model,
and it is for this reason we feel this approach will prove very useful
in studying the much more complicated electroweak model at high
temperature. In this paper we denote the 3d scalars
$\varphi$ and the 4d scalars $\phi$. The 4d gauge and scalar
couplings are $g$ and $\lambda$, respectively.

For $6g^2/\lambda\leq O(1)$,
by introducing an auxiliary field $\chi$ we  eliminate the
explicit $\varphi^4$ term and develop a systematic $1/N$ expansion for
$V_{eff}$  and
compute the leading and next--to--leading corrections in this
expansion.
By matching linearly divergent integrals in the effective 3d
theory with the $O(T^2)$ corrections in the 4d theory we show
how to obtain exactly the correct high temperature result of the
full 4d model, at least in the $g=0$ case which is known [6].
 The effective 3d model
is sufficient to study the nature of the
phase transition which occurs at values of $\phi^2$ and $\chi$
much less than $T^2$.  In contrast to previous studies,
we find no $|\phi|$ term in $V_{eff}$, nor
for $6g^2/\lambda\ll N$ any significant $|\phi|^3$ term in $V_{eff}$.
In fact, to the order we compute, the model
admits only a second order phase transition. First order phase
transitions are however found for $6g^2/\lambda\sim O(N)$ and
$6g^2/\lambda\sim O(N^{2\over3})$.

We mention that
the high temperature 3+1 dimensional
abelian Higgs model has also been investigated in [7] as an effective
3 dimensional theory. However, in that reference, the
$\epsilon$--expansion method is used to study the nature of the
phase transition and it does not shed any light on the above
questions and problems.

In section 2 we show how the pure scalar result can be
obtained in our approach and find agreement with an earlier
calculation of this author [6].
In section 3 we extend the analysis
to the gauged case for $6g^2/\lambda\leq O(1)$. In section
4 we find the leading contributions to $V_{eff}$ for $6g^2/\lambda
\sim O(N)$ and $O(N^{2\over3})$. In section 5 we discuss our results
and the phase transition and our conclusions
appear in section 6.
The pure scalar example, although previously
studied at finite temperature [6], sets much of the foundation
for the analysis in section 3, including in particular the
vacuum polarization effects.

\newpage
\noindent {\Large \bf 2. $\phi^4$ Model to Subleading Order.}

First consider a model with $N$ real scalars $\phi^i$ with the
tree level 3+1 dimensional Lagrangian
\be L[\phi]= {N\over2}\delta_{ij} \partial_\mu\phi^i\partial^\mu
  \phi^j-{N \lambda\over 4!}(\phi^2-v^2)^2. \ee
$\lambda$ is a dimensionless coupling and $v$ is the tree level
$vev$ of $|\phi|$. The space--time metric is
$\eta^{\mu\nu}={\rm diag}(+,-,-,-)$. Furthermore,
to avoid problems associated with triviality we will
assume $\lambda>0$ and consider the model as an effective low energy
model valid below some scale $\bar{\Lambda}$ [10].

The systematic $1/N$ expansion allows us to calculate (1) as
a perturbation in $1/N$ near $\phi=0$ [8,9]. Root [9] has evaluated
the leading corrections and given formal expressions for the
next--to--leading corrections to the
zero temperature scalar potential in 4,3,2 and 1 dimensions. The
procedure at finite temperature is very similar.

\def\l{\lambda}
By introducing a dimension
two auxiliary field $\chi$ we replace the Lagrangian (2):
\begin{eqnarray} L[\phi,\chi] &=& L[\phi] + {3N\over 2\l}
   \(\chi-{\l\over 6}(\phi^2-v^2)\)^2
   \nonumber  \\ & = &
   {N\over2} (\partial\phi)^2+  {3N\over 2\l} \chi^2
      -{N\over2}  (\phi^2-v^2)\chi.
\end{eqnarray}
The auxiliary field has eliminated the $\phi^4$ term;
the original form of (3) is easily recovered by use of the
equation of motion for $\chi$.

To calculate the effective potential $V_{eff}(\phi)$ one proceeds as
follows. First, using the background field method one computes the
effective potential as a function of backgrounds of $\phi$ and
$\chi$. Then, the background of $\chi$ is
eliminated by its equation of motion.

The systematic $1/N$ expansion first requires expanding the Lagrangian
$L[\phi,\sigma]$ about real backgrounds $\phi$ and $\chi$ thus [8,9,11]:
\be \phi^i \rightarrow \phi^i+{\hphi^i\over\sqrt{N}}, \qquad
     \chi \rightarrow \chi+ {\hat{\chi}\over\sqrt{N}}, \ee
and then deleting terms linear in the quantum fields $\hphi$,
$\hat{\chi}$. This procedure defines a quantum Lagrangian
$L[\hphi,\hat{\chi}]$, the sum of
whose one--particle irreducible (1PI) diagrams give what we call
the effective action.
Since we are interested in the effective
potential we assume the backgrounds are space--time constants.
We also assume non-negative $\chi$.

At finite temperature the model is formally equivalent to a
euclidean field theory with one compact dimension.
When the backgrounds $\phi^2$
and $\chi$ are much below $T^2$ it is sufficient to study an
effective three dimensional theory, a fact that we exploit below.
Thus, at sufficiently
high temperature, our results for four dimensions
should be similar to  those of a three dimensional euclidean field theory
with a dimensionful $\phi^4$ coupling [7]. Indeed,
in three dimensions and zero
temperature, the leading $O(N)$ potential has long been known [8].
It has exactly the same form as the sum of finite temperature superdaisy
graphs that were computed by Dolan and Jackiw [1] for a four
dimensional $\phi^4$ theory.
An important point in our approach is that
the effect of introducing an auxiliary field
$\chi$ is to shift the $\phi$ mass term and  as a result there are
no infrared divergences in this formalism.

\def\hchi{\hat{\chi}}\def\vx{\vec{x}}\def\vd{\vec{\partial}^2}
To perform the dimensional reduction from 4 to 3 dimensions, one
can follow [7]. One decomposes the fields as follows,
\be \hphi^i(x^\mu)=\sum \hphi^i_n(\vx)\psi_n(\tau),\qquad
    \hchi(x^\mu)=\sum \hchi_n(\vx)\psi_n(\tau),
\ee
where $\psi_n(\tau)$ are a complete set of periodic functions
on the circle. For what we are interested in only the zero modes $n=0$,
for which
$\psi_0(\tau)=\psi_0(0)$, are important in the effective 3d model.
This is because for $n\neq 0$ the fields $\hphi_n^i$ pick up nonzero
masses of $O(T^2)$ when the compact dimension is integrated out in the
action. Therefore, truncating the spectrum to keep only the
zero modes and integrating out the compact dimension gives
the effective 3d euclidean Lagrangian ($\beta=1/T$)
\begin{eqnarray} {1\over\beta} L[\hphi_0,\hchi_0] & = &
   -{3N\over 2\l}\chi^2 + {N\over2} \chi (\phi^2-v^2)
  \nonumber \\ & &
   +{1\over2}\hphi^i_0 [\delta_{ij}(-\vd+\chi)]
    \hphi^j_0+{\hphi^2_0\hchi_0\over 2 \sqrt{N}}
    +\phi^i\hphi^i_0\hchi_0
    -{3 N\hchi^2_0\over 2\l} . \end{eqnarray}
Recall that $\phi$ and $\chi$ are space--time constants.
Defining the three dimensional quantities,
\def\vp{\varphi}\def\hvp{\hat{\varphi}}
\def\tl{\tilde{\lambda}}\def\tv{\tilde{v}}
\be \vp^i = \sqrt{\beta}\phi^i,\quad \hvp^i=\sqrt{\beta}\hphi^i_0,
  \quad \tl = \l/\beta, \quad \tv^2 =\beta v^2,
\ee
we finally obtain the 3d quantum Lagrangian
\def\hc{\hchi}\def\ch{\chi}
\begin{eqnarray}  L[\hvp,\hc] & = &
  -{3N\over 2\tl}\ch^2 + {N\over2} \ch (\vp^2-\tv^2)
  \nonumber \\ & &
   +{1\over2}\hvp^i [\delta_{ij}(-\vd+\ch)]
    \hvp^j+{\hvp^2\hc\over 2 \sqrt{N}}
    +\vp^i\hvp^i\hc
    -{3 N \hc^2 \over 2\tl} . \end{eqnarray}
Here we have dropped the zero  mode subscript on $\hchi_0$.

Before proceeding we would like to stress some important points.
As shown in [7], if one integrates out the nonzero modes $n\neq 0$
at the quantum level rather than just truncating the spectrum, there
is a finite $O(T^2)$ correction to the mass of $\hphi^i_0$. This
correction is very important, in fact it is the term that gives
symmetry restoration at high enough temperature. We will obtain
this term another way, in analogy with what one does in effective
low energy theories of the strongly interacting standard model
or effective four dimensional supergravity models inspired by
string theory [12]. The three dimensional field theory will be
divergent. Here
we will regulate all the divergent integrals by introducing
the same scale $\Lambda$. In fact, the only divergent integral that will
be important is linearly divergent and we will regulate it by
simply using a sharp momentum cutoff.\footnote{A different scheme,
such as Pauli--Villars, will in general give a different coefficient
for the $O(\Lambda)$ result. However, the precise coefficient
will not be important in what follows [12].}
We then give a physical interpretation
to $\Lambda$, i.e. the scale at which the full four dimensional
physics becomes important. Thus, $\Lambda$ is proportional to $T$.
This is in complete analogy with, for example, the effective
4d theories where the regulating scale is taken to be of order
the compactification scale. In our case the identification can be
made precise because the corresponding 4d correction is well
known. We will see that the identification $\Lambda=\pi^2 T/6$
will reproduce exactly the $T^2$ results from the four dimensional
integrals.
In addition, since the four dimensional model (2) is
only valid up to a scale $\bar{\Lambda}$ we must require
$\Lambda<\bar{\Lambda}$, i.e. that $T$ is sufficiently small.
Finally we note that the 4d effective potential can be obtained
from the three dimensional one by dividing by $\beta$ and using (7).

To compute the 3d effective potential we must sum the
one--particle irreducible (1PI) diagrams of (8). For readability, and
for later use in the abelian Higgs model, we briefly describe the
steps that lead to the next--to--leading result for $V_{eff}$, rather
than just stating the result.
Throughout, Tr stands for momentum and internal
space traces.

Integrating out the  $\hphi^i$
in this model gives the $O(N)$ corrections to $V_{eff}$ to all
orders in $\hbar$ and defines an effective Lagrangian which only has
quantum dependence on $\hat{\chi}$. The part of this effective Lagrangian
which is quadratic in $\hat{\chi}$ includes the tree level part and
a one--loop vacuum polarization part.
The one--loop $\hat{\chi}$
corrections give the next--to--leading corrections in $N$, also to
all orders in $\hbar$.

Because (8) is quadratic in $\hvp$, the $O(\hvp\hc)$ mixing term
can be removed by making a change of field variables [6]. This
has the effect of replacing the mixing term by
\be +\vp^i\hvp^i\hc\rightarrow  -{\hc^2\over2}{\vp^2\over
   (-\vd+\chi+\hc/\sqrt{N})} \rightarrow -
   {\hc^2\over2}{\vp^2\over (-\vd+\chi)}, \ee
where the last replacement holds to next--to--leading order in
the $1/N$ expansion.

\def\old{ \begin{picture}(77,22)(0,+8.5)
     \thicklines
  \put(4,10){\line(1,0){22}}\put(4,12){\line(1,0){22}}
  \put(36,11){\circle{20}}
  \put(46,10){\line(1,0){23}}\put(46,12){\line(1,0){23}}
\end{picture} }
The gaussian integral over $\hvp$ is now straightforward
and its vacuum polarization contribution to the $\hc$
propogator at nonzero external momentum is known [8,9] in 3d,
\be \Pi_{\hc \hc }(\vec{k})= \old =
{\tl\over 6} \int_{\vec{p}}
   {1\over [ \vec{p}^2+\chi] [  (\vec{p}+\vec{k})^2+\chi]  } =
   {\tl\over 24\pi}{1\over\sqrt{\vec{k}^2}}\sin^{-1}
   \({1\over\sqrt{1+4\chi/ \vec{k}^2}}\). \ee
The final result for the leading + next--to--leading potential is [8,9,6]
\be V_{eff}=
     - {3N\over2\tl}\chi^2+{N\over2}\chi
    (\vp^2-\tv^2)+{N\over2}\Tr\ln (-\vd+\chi) + V_{Ntl},
\ee
where the next--to--leading contribution from the gaussian
integral over $\hat{\chi}$ is
\be V_{Ntl}= {1\over2}\Tr\ln
   \[ 1+{\tl\over 24\pi}{1\over\sqrt{-\vd}}\sin^{-1}
   \({1\over\sqrt{1-4\chi/ \vd}}\)
           + {\tl\vp^2/3 \over -\vd+\chi}
    \]. \label{enl}
\ee

Root [9] has shown, without explicit calculation, that the
3d field theory is renormalizable to this order in $N$, as it
should be. Although we will give a physical interpretation to
our regulating scale, renormalizability of the 3d theory places
strong constraints on the type of linearly divergent contributions
we can obtain. For example, a contribution of $O(\Lambda \vp^2)$
is by itself not renormalizable in our formalism. How this enters
will be seen in the explicit calculations below.

Although we did not find a simple expression for $V_{Ntl}$, we found that
the dominant contribution comes from the large external momentum
limit of (10), $-\vd\gg \chi$,
$not$ the zero external momentum limit. Although
this seems entirely reasonable in the limit of vanishing $\chi$,
why it should be so for sufficiently large $\chi$ is not apparent.
Indeed, we would like to stress
the importance of vacuum polarization effects at nonzero
external momentum. If in (12) we had kept only the zero external
momentum part of the $\hc$ field vacuum polarization,
$\Pi_{\hc\hc}(0)=\tl/(48\pi\sqrt{\chi})$ then we would have obtained
an incorrect (and nonrenormalizable) answer. $\Pi_{\hc\hc}(0)$
when used for the whole momentum integral from 0 to $\Lambda$
gives a dangerous $O(\tl\Lambda^3/\sqrt{\chi})$ contribution
in the expansion of the log in (12). In fact, the leading field
dependent contributions seem to arise from the most ultraviolet
divergent field dependent term in the high momentum expansion of
$\Pi_{\hc\hc}$. One can check this assertion when $\chi$
is sufficiently large by the following method. In this limit, the
arcsine term in (12) is small for all values of the momentum, and
we expand the $\ln$ using $\ln[1+x]=x+...$. The arcsine possesses
two different expansions depending on whether $-\vd$ is bigger
or smaller than $4\chi$, and accordingly the momentum integral
implied by the Tr must be broken up into two different regions.
The different contributions can then be evaluated with the result
that very little error is made is using only the large momentum
limit of (10) everywhere.
This effect will also be very important in
the next section.

In the large momentum limit, we have
\be  (-\vd+\chi)\Pi_{\hc\hc}(-\vd) \rightarrow
      {\tl\sqrt{-\vd}\over 48} - {\tl\sqrt{\chi}\over12\pi}. \ee
We then write eq. (12) as
\be
V_{Ntl} \approx {1\over2}\Tr\ln
   \[ -\vd+\chi +  {\tl\over 48} \sqrt{-\vd} +
            {\tl\over3}\(\vp^2-{\sqrt{\chi}\over 4\pi}\)\]
    -{1\over2}\Tr\ln\[-\vd+\chi\].
\ee
It can now be argued that if $\tl/48 \ll \Lambda$ and $\chi$
is sufficiently large the $O(\sqrt{-\vd})$ term in the log
can be dropped [13]. This is what we will assume.

We can now give an explicit expression for $V_{eff}$, up to
the approximations we have made by using the following result
for three dimensions ($\chi\geq 0$):
\be \Tr\ln(-\vd+\chi)\rightarrow \int {d^3\vec{k}\over (2\pi)^3}
   \ln(\vec{k}^2+\chi) =
{\Lambda\chi\over 2\pi^2}-{\chi^{3\over2}
   \over 6\pi} + {\rm const.} \ee
A sharp momentum cutoff was used to evaluate the field dependent
divergent part. Adding everything up we obtain, up to a constant,
\begin{eqnarray} V_{eff}&=&
     - {3N\over2\tl}\chi^2+ {N\over 2}\chi
    (\vp^2-\tv^2)+ N\[ {\Lambda\chi\over 4\pi^2}-{\chi^{3\over2}
    \over 12\pi} \] +{\Lambda\tl\over 12\pi^2}\(\vp^2-{\sqrt{\chi}
    \over 4\pi}\) \nonumber \\ & & -{1\over12\pi}\(
   \(\chi+{\tl\over3}\vp^2-{\tl\sqrt{\chi}\over 12\pi}\)^{3\over2}
     -\chi^{3\over2} \) .
\end{eqnarray}
To write the effective potential for $\vp^2$ alone we use the
equations of motion for $\chi$ to eliminate it. To the order
we are working  it is consistent to use the leading order equation
for $\chi$ to eliminate it [9]. $\partial V_{eff}/\partial\chi=0$ from
the terms proportional to $N$ gives:
\be \chi =  {\tl\over6}\(\vp^2-\tv^2\)+{\tl\over 6}\(
      {\Lambda\over 2\pi^2}
        -{\sqrt{\chi}\over 4\pi}\) . \label{mge}
\ee

If we rewrite the $O(1)$ terms in $V_{eff}$ using this equation,
and ignore $\tl/(12\pi)$ in comparison with $\sqrt{\chi}$ then
we obtain up to a constant,
\be V_{eff} =
     - {3N\over2\tl}\chi^2+ {N\over 2}\chi
    (\vp^2-\tv^2)+ (N+2) {\Lambda\chi\over 4\pi^2}
     -{1\over12\pi}\(  \(\chi+{\tl\over3}\vp^2\)^{3\over2}
     +(N-1) \chi^{3\over2} \) . \label{vs}
\ee

This is the leading + next--to--leading result for $V_{eff}$ for
a pure scalar theory in 3d when $\sqrt{\chi}$ is large enough.
To obtain the high temperature four dimensional result, we recall
the high temperature 4d leading order mass--gap equation [1,6]:
\be \chi={\l\over6} (\phi^2-v^2)+{\l\over6}\({1\over 12\beta^2}
    - {\sqrt{\chi}\over4\pi\beta}\)+\ldots . \label{fr}\ee
Using (7), we see that the identification
\be \Lambda=\pi^2 T/ 6 \ee
in (\ref{mge}) reproduces exactly
the four dimensional result (\ref{fr}). In fact,
with this identification, (\ref{vs}) reproduces exactly the
leading $and$ next--to--leading 4d
finite temperature scalar potential found in [6]. The computations
here have been much easier. This potential was studied in [6] and
admits no first order
phase transition for the range of $\phi^2$ and $T$ where it is valid
(i.e. not too close to the origin when the temperature is near
$T_2$). Similar conclusions appear in [7,14].
Our result is essentially due to the fact that $V_{eff}$ contains
no $O(|\phi|^3)$ terms for $\l\phi^2<3\chi$.

The potential (18) is valid when $\tl/(12\pi)$ is ignorable
compared to $\sqrt{\chi}$.
We will give a more extensive discussion of the phase transition
in the more general gauged case,
in the last section.

\newpage
\noindent {\Large \bf 3. Abelian Higgs Model, $6g^2/\l\leq O(1)$.}

Now consider the 3+1 dimensional gauge invariant
scalar QED Lagrangian
\be L[\phi,A]= {N\over2}\delta_{ij} \partial_\mu\phi^i\partial^\mu
  \phi^j-{N \lambda\over 4!}(\phi^2-v^2)^2
   -{1\over 4} F^{\mu\nu}F_{\mu\nu}-g\epsilon_{ij}(\partial_\mu\phi^i)
   \phi^j A^\mu + {1\over2} g^2 \phi^2 A^2 .\ee
Here, $g$ is the gauge coupling, $\phi^i$, $i=1...N$, are real
as before and $\epsilon_{ij}$ is antisymmetric with the nonzero
components $\epsilon_{12},\epsilon_{21},\epsilon_{34},\epsilon_{43},
...$ all having magnitude 1. The field strength and covariant
derivative $D_\mu$ are given in terms of the gauge field $A^\mu$ by
\begin{eqnarray}
   D_\mu\phi^i &=&
   \partial_{\mu}\phi^i-g\epsilon_{ij}\phi^j A^\mu, \nonumber \\
  F_{\mu\nu} &=& \partial_{\mu} A_\nu-\partial_\nu A_\mu .
\end{eqnarray}
This model is nothing but $N/2$ copies of the simple scalar QED
Lagrangian discussed in many textbooks, and also in ref. [1], but
with only one gauge vector field.

We would like to calculate the leading + next--to--leading
corrections to $V_{eff}$ in the $1/N$ expansion at high temperature.
As in section 2, we will do this by first writing an effective
3d theory. This three dimensional theory will turn out to be an
abelian Higgs model with an extra scalar. This can be understood
intuitively because in 4 dimensions the gauge field has three
massive components, while in three dimensions it only has two. Hence,
to get the same light degrees of freedom we need an additional scalar.

To proceed, we introduce an auxiliary field, as in eq. (3), to
obtain
\be L[\phi,\chi,A] =
   {N\over2} (\partial\phi)^2+  {3N\over 2\l} \chi^2
      -{N\over2}  (\phi^2-v^2)\chi
   -{1\over 4} F^{\mu\nu}F_{\mu\nu}-g\epsilon_{ij}(\partial_\mu\phi^i)
   \phi^j A^\mu + {1\over2} g^2 \phi^2 A^2 .\ee
We then expand this using (4), and delete all terms linear in
$\hphi$ and $\hc$. Since we are interested in the scalar potential
we need not keep a background for the gauge fields.

We also need to gauge fix. We add the gauge fixing term
\be L_{g.f.}= -{1\over 2\alpha}(\partial_\mu A^\mu
   +\alpha g \epsilon_{ij} \hphi^i \phi^j)^2 . \ee
Here, $\alpha$ is an arbitrary parameter. Although the calculations
simplify in the Landau gauge $\alpha\rightarrow 0$, this extra
parameter is useful in checking that physical quantities such
as the critical temperature are gauge fixing independent.
The ghost term for this gauge fixing is [1]
\be L_{gh}= \bar{\theta} (\partial^2+\alpha g^2\phi^2)\theta, \ee
where $\theta$, $\bar{\theta}$ are Grassmanian
ghost fields. We assume non-negative
$g^2,\alpha$ and $\alpha\leq O(1)$ [4].

After performing all these operations, the total quantum Lagrangian
is given by
\begin{eqnarray}
 L[\hphi,\hc,A,\theta] & = &  L[\hphi,\hc]
  +{1\over 2} A_\nu \partial^2 A^\nu + {1\over2}(1-{1\over\alpha})
   (\partial A)^2\nonumber \\
 & & +{1\over2} g^2 (\phi^2+2\phi_i\hphi^i/\sqrt{N}+\hphi^2/N)A^2
   -{g\over\sqrt{N}}\epsilon_{ij}(\partial_\mu\hphi^i)\hphi^j A^\mu
   \nonumber \\
 & & -{1\over2}\alpha g^2\hphi^i\hphi^k\epsilon_{ij}\epsilon_{kl}
   \phi^j\phi^l+\bar{\theta}(\partial^2+\alpha g^2\phi^2)\theta .
\end{eqnarray}
We dropped all total divergences and assumed the backgrounds are
space--time constants. There is no $O(\hphi A)$ term because of
our gauge choice. Finally, $L[\hphi,\hc]$ is the pure scalar
part of the quantum Lagrangian.

To write the 3d effective field theory for the high temperature
limit of this model  we follow exactly the
approach described in section two. We simply truncate the
spectrum to keep the zero modes of the compact dimension
of the euclidean theory
and then give a physical interpretation to the scale used to
regulate divergent integrals. This truncation means
$\partial_0\rightarrow 0$. Then, with the definitions (7)
as well as ($a$=1,2,3)
\def\A{{\cal A}}\def\tg{\tilde{g}}\def\vt{\vartheta}
\be \A^a=\sqrt{\beta}A^a,\quad \rho=i\sqrt{\beta}A^0,\quad
   \vt=\sqrt{\beta}\theta,\quad \tg = g/\sqrt{\beta}, \ee
we write the effective 3d quatum Lagrangian as
\begin{eqnarray}
 L[\hvp,\hc,\A,\rho,\vartheta] & = &  L[\hvp,\hc]
  -{1\over 2} \A_a \vd \A^a - {1\over2}(1-{1\over\alpha})
   (\vec{\partial} \A)^2- {1\over2}\rho\vd\rho\nonumber \\
 & & +{1\over2} \tg^2 (\vp^2+2\vp_i\hvp^i/\sqrt{N}+\hvp^2/N)
      (\A^2+\rho^2)
   -{\tg\over\sqrt{N}}\epsilon_{ij}(\partial_a\hvp^i)\hvp^j \A^a
   \nonumber \\
 & & +{1\over2}\alpha \tg^2\hvp^i\hvp^k\epsilon_{ij}\epsilon_{kl}
   \vp^j\vp^l+\bar{\vt}(-\vd+\alpha \tg^2\vp^2)\vt. \label{aht}
\end{eqnarray}
$L[\hvp,\hc]$ is given by eq. (8), and $\A_a=\A^a$, $\partial_a
=\partial^a$.

Eq. (\ref{aht}) is nothing but a three dimensional abelian Higgs model
with an extra massless scalar field $\rho$.
It has a simple canonical kinetic
term and its tree level coupling to the scalars is simply
$+\tg^2\vp^2\rho^2/2$.
We now compute the 1PI diagrams of this 3d Lagrangian to leading
and next--to--leading order in the $1/N$ expansion. It is clear
that the gauge sector contributes only at next--to--leading order;
there are no Feynman diagrams involving the gauge fields that
contribute at $O(N)$. The correction from the 3d ghosts involves
only a simple one--loop calculation. For the rest, we can either
first integrate out the $\A$ and $\rho$ at the quantum level,
or integrate out the $\hvp$. We have done both, and present only
the latter computation.

To integrate out the $\hvp$ we must first make a field
redefinition to eliminate terms in (28) which are linear in
$\hvp$. For $\tg=0$ this is just the replacement (9) to the
required order. For $\tg\neq 0$ we have to shift $\hvp$ in
such a way as to also eliminate the $O(\tg^2)$ linear terms in $\hvp$.
When this is done we found no contributions from the $O(\tg^2\hvp^i)$
terms that were important at next--to--leading order. Another way
of saying this is that there are no Feynman diagrams involving
the $O(\tg^2\hvp^i)$ terms that contribute at next--to--leading order.
Hence we just forget these terms in the Lagrangian.

In the pure scalar case we saw that integrating out the scalars
gave the leading order result and generated a vacuum polarization
term $\Pi_{\hc\hc}$
for $\hc$. Here, something similar happens. In addition to
$\Pi_{\hc\hc}$ we will generate a vacuum polarization $\Pi_{\rho\rho}$
for $\rho$ and a vacuum polarization matrix $\Pi^{ab}$ for the
3d gauge fields.
This result can be derived very simply as follows. The scalar loop
integral in (28) contributes a piece
\be {1\over 2} \Tr\ln\[ \Delta^{-1}_{ij}+\delta_{ij}{\hc\over\sqrt{N}}
+\delta_{ij}{\tg^2\over N}
   (\A^2+\rho^2)-2 {\tg\over \sqrt{N}}\epsilon_{ji}\A_a\partial^a\]
   \label{sc} \ee
where the inverse scalar propogator is given by the matrix
\be \Delta^{-1}_{ij}=\delta_{ij}(-\vd+\chi) + \alpha\tg^2
   \epsilon_{ii'}\epsilon_{jj'}\vp^i\vp^{j'} .\ee
One can expand (\ref{sc}) as a power series in $\A$; the terms
linear in $\A$ in this expansion vanish because tr$\epsilon_{ij}=0$
and only the terms quadratic a $\A$ and $\rho$
 contribute at $O(1)$ in the $1/N$
expansion using tr$\epsilon_{ij}\epsilon_{jk}=-N$. At zero
external momentum the expansion is easy and gives to $O(1)$
\be {1\over2}\Tr\ln(\Delta^{-1}_{ij}+\delta_{ij}\hc/\sqrt{N})
     +{1\over2}\tg^2(\A^2+\rho^2)
  \Tr (-\vd+\chi)^{-1} + \tg^2\A_a\A_b\Tr \partial^a\partial^b/
     (-\vd+\chi)^2 .\ee
The $\A^2,\rho^2$ terms are clearly vacuum polarization corrections
to the tree level kinetic terms of these fields. The vacuum polarization
for $\hc$ is contained in the first term above, and is in fact the
same as in the pure scalar case at $O(1)$. Since we have seen that
vacuum polarization effects at nonzero external momentum are
very important even in the pure scalar case we will also keep
them here. This can be done by carefully accounting for the
3-space dependence of all the quantum fields in (\ref{sc}) and
the result for the full $\A,\rho$ vacuum polarizations is
nothing other than the one--loop result given for scalar QED in many
standard textbooks [15]. Altogether, one has
\def\ola{ \begin{picture}(71,22)(0,+8.5)
     \thicklines
  \multiput(4,11)(8,0){3}{\oval(4,4)[b]}
  \multiput(8,11)(8,0){3}{\oval(4,4)[t]}
\put(36,11){\circle{20}}
  \multiput(48,11)(8,0){3}{\oval(4,4)[b]}
  \multiput(52,11)(8,0){3}{\oval(4,4)[t]}
\end{picture} }
\def\olb{ \begin{picture}(59,22)(0,+8.5)
     \thicklines
  \multiput(4, 5)(8,0){7}{\oval(4,4)[b]}
  \multiput(8, 6)(8,0){7}{\oval(4,4)[t]}
\put(32,18){\circle{20}}
\end{picture} }
\def\olc{ \begin{picture}(77,22)(0,+8.5)
     \thicklines
  \multiput(4,11)(8,0){3}{\line(1,0){6}}
\put(36,11){\circle{20}}
  \multiput(46,11)(8,0){3}{\line(1,0){6}}
\end{picture} }
\begin{eqnarray}
  \Pi_{\hc\hc}(\vec{k}) & = & \old =
{\tl\over 6} \int_{\vec{p}}
   {1\over [ \vec{p}^2+\chi] [  (\vec{p}+\vec{k})^2+\chi]  },
\nonumber \\
  \Pi_{\rho\rho}(\vec{k}) &=& \olc = \tg^2\int_{\vec{p}}{1\over
       \vec{p}^2+\chi} ,
 \nonumber \\
  \Pi^{ab}(\vec{k}) &=& \ola + \olb = \tg^2\int_{\vec{p}}\[
  {\delta^{ab}\over \vec{p}^2+\chi} -
     {{1\over2} (2p+k)^a(2p+k)^b
         \over[\vec{p}^2+\chi][(\vec{p}+\vec{k})^2
        +\chi]}\] .
  \label{vpe}
\end{eqnarray}

$V_{eff}$ then is given by the tree--level piece, a contribution
from the scalars loops, ${1\over2}\Tr\ln\Delta^{-1}_{ij}$, a simple
one--loop contribution from ghosts $\vartheta$, and the contributions
from quadratic integrals over $\A$, $\rho$, $\hc$ with propogators
modified by the vacuum polarization effects(\ref{vpe}). The result
is:
\begin{eqnarray}
  V_{eff} &=&
     - {3N\over2\tl}\chi^2+{N\over2}\chi
    (\vp^2-\tv^2)+{1\over2}\Tr\ln \Delta^{-1}_{ij} + V_{Ntl} \nonumber\\
   & & - \Tr\ln(-\vd+\alpha \tg^2 \vp^2) + {1\over2}\Tr\ln(-\vd
       +\tg^2\vp^2+\Pi_{\rho\rho}) +{1\over2}\Tr\ln(
        \Delta^{-1}_{ab}+\Pi_{ab}).
\end{eqnarray}
The tree--level gauge kinetic term is given by
\be \Delta^{-1}_{ab} = -\delta_{ab}\vd+(1-\alpha^{-1})\partial_a
    \partial_b+\tg^2\vp^2\delta_{ab} .\ee
The contribution from $\hc$ is the same as in the pure scalar case,
eq. (12), because $\Pi_{\hc\hc}$ did not change to $O(1)$ in $1/N$.

A few remarks are in order:\\ $\bullet$
 In the Landau gauge $\alpha\rightarrow 0$,
the 3d gauge contributions are given only by the last term in (33).
In this gauge, the $\rho$ and $\A_a$
contributions are similar to the 4d gauge loop
ring sum worked out in refs. [2,4]. As mentioned in the introduction,
their procedure overcounts the $O(\hbar^2)$ graphs.
For the $\tg$ dependent $O(\hbar^2)$
graphs the overcounting is due to an additional contribution
from the scalars
because for the scalar loop contribution they use propogators
with $\tg$ dependent vacuum polarization effects. Our method of
computing $V_{eff}$ automatically avoids this overcounting.
In addition our contribution from (33) includes
much more than the ring sum  in [2,4] because our vacuum polarization
effects depend on $\chi$ which is a
self consistent solution of $\partial V_{eff}/\partial\chi=0$ to all
orders in $\hbar$, whereas the vacuum polarization used in [2,4] is
of $O(\hbar)$ only. \\ $\bullet$
In a general gauge, $\alpha\neq 0$, the contributions to $V_{eff}$
due to nonzero $\Pi_{ab}$ are independent of $\alpha$. This is
a result due to gauge invariance, i.e.  $k^a \Pi_{ab}(\vec{k})=0$
formally holds for (32) -- as it also should for a properly regulated
$\Pi_{ab}$. We have $\Delta^{-1}_{ab}+\Pi_{ab}=\delta_{ab}(-\vd
+\tg^2\vp^2)+[(1-\alpha^{-1})\partial_a\partial_b+\Pi_{ab}]$. The
log of this can formally be expanded as a power series in the
term in square brackets. Since $\Pi_{ab}$ is transverse,
tr$[(1-\alpha^{-1})\partial_a\partial_b+\Pi_{ab}]^n$ factorizes
so that there are no mixed $\alpha$--dependent and $\Pi$ dependent terms.
Altogether, this means
\begin{eqnarray}
 \Tr\ln[\Delta^{-1}_{ab}+\Pi_{ab}]&=&\Tr\ln\Delta_{ab}^{-1}
   +2\Tr\ln(-\vd+\tg^2\vp^2+\Pi)-2\Tr\ln(-\vd+\tg^2\vp^2) \nonumber \\
 &=& \Tr\ln(-\vd+\alpha\tg^2\vp^2)+2\Tr\ln(-\vd
       +\tg^2\vp^2+\Pi). \end{eqnarray}
Here, $\Pi_{ab}=(\delta_{ab}-k_a k_b/\vec{k}^2)\Pi$ which is
true due to $k^a\Pi_{ab}=0$ and 3d euclidean invariance.
\\ $\bullet$
Renormalizability of the model places strong contraints on the type
of linearly divergent corrections to $V_{eff}$. The reason is the
same as in the pure scalar case and  as in section 2 we cannot
obtain, by itself, a $O(\tl\Lambda \vp^2)$ contribution to $V_{eff}$.
\\ $\bullet$ As already described we will give a physical
interpretation to the regulating scale for divergent integrals.
In general, one should introduce arbitrary parameters for the
different divergent terms [12]. In the pure scalar case we did
not do this because it turned out to be unnecessary. In our
computations for this section it turned out that the match between
the linearly divergent 3d integrals and the finite $T^2$ corrections
of the full theory required that the linearly divergent part
of $\Pi_{\rho\rho}$ be regulated not with $\Lambda\rightarrow\pi^2 T/6$
but with $\Lambda'\rightarrow\pi^2 T/3$. In general, we should also
use unspecified sharp momentum cutoffs for the
linear divergences
from the final $\rho$ and $\A_a$ loop contributions. However, the same
cutoff $\Lambda\rightarrow \pi^2 T/6$ that gave the $O(T^2)$
corrections in the scalar case also worked here.
This should be  kept in mind in what follows.

Many of the traces in (34) are straightforward to compute. Using
(15) we find  up to field independent terms,
\begin{eqnarray} \Tr\ln\Delta_{ij}^{-1} &=&
   N\[ {\Lambda\chi\over2\pi^2}-{\chi^{3\over2}\over 6\pi}\]
     +{\alpha \tg^2 \Lambda\vp^2\over 2\pi^2}
     -{1\over 6\pi}\[ (\chi+\alpha\tg^2\vp^2)^{3\over2}-\chi^{3\over2}
        \], \nonumber \\
 \Tr\ln(-\vd+\alpha\tg^2\vp^2) &=&
     {\alpha \tg^2 \Lambda\vp^2\over 2\pi^2}
     -{\alpha^{3\over2}\tg^3|\vp|^3\over 6\pi} , \nonumber \\
 \Tr\ln(-\vd+\tg^2\vp^2+\Pi_{\rho\rho}) &=&
     {\tg^2 \Lambda(\vp^2-{\sqrt{\chi}\over4\pi}) \over 2\pi^2}
     -{\tg^3\({\Lambda'\over 2\pi^2}+
\vp^2-{\sqrt{\chi}\over 4\pi}\)^{3\over2}
    \over 6\pi} . \end{eqnarray}
Notice that the contribution from the 3d scalar $\rho$ occurs
only in the combination $\vp^2-\sqrt{\chi}/4\pi$, a fact that is demanded
by multiplicative
renormalizability of the 3d theory.\footnote{Note that
due to our aim of
calculating the high temperature $V_{eff}$,  we identify
$\rho$ with the longitudinal 4d gauge field so that a bare mass
for $\rho$ is neither necessary nor allowed.}

We cannot compute the last term in (33) without first evaluating
$\Pi_{ab}$ at nonzero external momentum. This can be explicitly
done by following [15]. Gauge invariance
forbids a linearly divergent gauge mass correction and in dimensional
regularization the vacuum polarization displays no poles at
$d=3$\footnote{In 3d this last statement
can be understood as the requirement
that there should be no nonlocal ultraviolet counterterms [16].}.
Since $\Pi_{ab}=(\delta_{ab}-k_a k_b/\vec{k}^2)\Pi$,
we simply compute $\Pi={1\over2} \delta^{ab}\Pi_{ab}$. We find
\be \Pi ={\tg^2\over 8\pi}\[ -\sqrt{\chi} + {2\chi+{1\over2}\vec{k}^2
   \over \sqrt{\vec{k}^2}}\sin^{-1}
   \({1\over\sqrt{1+4\chi/ \vec{k}^2}}\) \]. \ee
This has the zero momentum value
\be \Pi =    0, \ee
and the high momentum limit
\be \Pi \sim -{\tg^2\sqrt{\chi}\over 4\pi}+{\tg^2\sqrt{\vec{k}^2}
   \over 32}. \ee
In the zero momentum limit $\Pi_{ab}=0$\footnote{This agrees
with the results given for the transverse components of the photon
in [5].}, and if this is wrongly
used in the whole momentum integral in (33) we will obtain a
nonrenormalizable answer for the 3d potential. This is most easily seen
in the Landau gauge for which all linearly divergent corrections
from the 3d scalar sector come in a combination proportional
to $\vp^2-{\sqrt{\chi}\over4\pi}$. The 3d gauge loop contributions for
$\Pi_{ab}=0$ however contribute $O(\tg^2\Lambda\vp^2)$ corrections
which are not renormalizable. Thus, keeping only the zero momentum
limit of $\Pi_{ab}$ is incorrect.

We believe that the dominant contributions to $V_{eff}$ from the
3d gauge loop integral arise from the $high$ momentum limit of
$\Pi_{ab}$; we  justify this as follows. For $\tg^2/12\pi$
ignorable in comparison to $\sqrt{\chi}$ it follows that
$\Pi(-\vd)$ is small compared to $-\vd$ for all values of $-\vd$.
We then expand $\Tr\ln(-\vd+\tg^2\vp^2+\Pi)$ as a power series
in $\Pi$. The momentum integrals implied by the $\Tr$ must be broken
up into the regions $-\vd>4\chi$ and $-\vd<4\ch$ in accordance
with the different expansions $\Pi$ possesses in these two
regions. At $O(\Pi)$ the different contributions can be computed
with the result that only a small error is made in using (39)
for the full momentum integral [17].

When $\Lambda\gg \vp^2-\sqrt{\chi}/4\pi \gg \tg^2/(64)^2$ [13]
we find, using eq. (35), a sharp momentum cutoff $\Lambda$ and ignoring
constants,
\be \Tr\ln(\Delta^{-1}_{ab}+\Pi_{ab})=
     {\alpha \tg^2 \Lambda\vp^2\over 2\pi^2}
     -{\alpha^{3\over2}\tg^3|\phi|^3\over 6\pi}
   + {\tg^2\(\vp^2-{\sqrt{\chi}\over4\pi}\) \Lambda\over \pi^2}
     -{\tg^3\(\vp^2-{\sqrt{\chi}\over4\pi}\)^{3\over2} \over 3\pi}.
\ee

Adding everything up, we obtain for the 3d $V_{eff}$,
\begin{eqnarray}
 V_{eff}&=& -{3N\over 2\tl}\chi^2+{N\over2}\chi(\vp^2-\tv^2)
     +{\Lambda\chi\over 4\pi^2}(N+2+{18\tg^2\over\tl})
     \nonumber \\ & &
    -{1\over12\pi}\(\(3\chi+{\tl\over3}(
        \tv^2-\Lambda/2\pi^2)\)^{3\over2}
    +(N-1)\chi^{3\over2} \) \nonumber \\
   & & -{(6\tg^2/\tl)^{3\over2}\over 12\pi}\[
     \(\chi+{\tl \tv^2\over6}+{\tl\over12\pi^2}(\Lambda'-\Lambda)
      \)^{3\over2}  + 2
     \(\chi+{\tl \tv^2\over6}-{\tl\over12\pi^2}\Lambda
      \)^{3\over2}\] \nonumber \\
  & & + V_\alpha +\; {\rm neglected\; terms} \; + O(1/N)\;{\rm terms}.
\end{eqnarray}
$V_{\alpha}$ is the pure gauge fixing dependent part,
\be V_\alpha=-{1\over12\pi}\[(\chi+\alpha\tg^2\vp^2)^{3\over2}
   -\chi^{3\over2}-\alpha^{3\over2}\tg^3 |\varphi|^3 \], \ee
and satisfies $V_\alpha(\chi=0)=V_\alpha(\vp^2=0)=0$. The ``neglected
terms'' in (41)
 refers to the (small) errors that were made in only keeping
the large external momentum limit of the vacuum polarizations, as well
as other approximations. Here, we also used the leading order
result for $\chi$, eq. (17),
to rewrite the next--to--leading terms in $V_{eff}$ --
a substitution that only changes the $O(1/N)$ terms.

To get the high temperature 4d $V_{eff}$ from (41)
we must match the $\Lambda$, $\Lambda'$ terms with the
4d result for $T^2$ finite pieces as well as multiplying
the 3d potential by $T$ and using eqs. (7), (27).
For the case $\tg=0$ we already
found $\Lambda=\pi^2T/6$ reproduces the correct 4d result. In fact,
this identification also works in the $\tg\neq 0$  case. This can
be checked by examining the $O(\hbar)$ (i.e. one loop in usual
perturbation theory) result for which the third term in (41) reduces
to [6] $\Lambda\vp^2(N\tl+2\tl+18\tg^2)/24\pi^2$. The identification
$\Lambda=\pi^2 T/6$ reproduces the one--loop result for scalar QED
given by Dolan and Jackiw [1]. Finally, $\Lambda'=\pi^2 T/3$ reproduces
the $O(T^2)$ scalar QED longitudinal gauge boson polarization
given in refs. [2,4,5]\footnote{To compare we must rescale $\vp^2
\rightarrow \vp^2/N$, $\tg^2\rightarrow \tg^2 N$ and set $N=2$.}.
To arrive at (41)  we made the assumptions
 $\Lambda\gg \sqrt{\chi}> \tl/12\pi,\tg^2/12\pi$,
$\Lambda\gg \vp^2-\sqrt{\chi}
/4\pi\gg \tg^2/(64)^2$ [13].
\newpage
\noindent{\Large\bf 4. Abelian Higgs Model, $6g^2/\l\sim O(N)$
 and $O(N^{2\over3})$.}

\noindent{\bf The case $6g^2/\l\sim O(N)$.} In section 3
we considered $\l$ fixed as $N$ increases. In this case we must
assume $(\lambda N)$ fixed as $N$ increases. For a weakly coupled
model this means the scalar self--coupling is very weak. First
let us consider the pure scalar case of section 2. We again work
in 3d. To obtain the leading corrections we make the
replacements $\tl=\tl'/N$, $\chi\rightarrow\chi/N$, $\hc\rightarrow
\hc/N$ in (8) and (11). Then $V_{Ntl}$ is at most of $O(1/N)$; the
remainder of (11) is of $O(1)$ and $O(1/\sqrt{N})$. Keeping only
the $O(1)$ and $O(1\sqrt{N})$ terms the answer can be
exactly computed. For non-negative $\chi$, (11) gives:
\be V_{eff}= -{3\over 2 N\tl}\chi^2 + {1\over 2}
   \chi (\vp^2-\tv^2)+ {\Lambda\chi\over 4\pi^2}-{\chi^{3\over2}
   \over 12\pi\sqrt{N}} + {\rm const.} \ee
$\chi$ should be eliminated by its leading order equation from
the the first three terms above. For $g\neq 0$ the computation
of all $O(1/\sqrt{N})$ terms is nontrivial. The replacements
$\tl=\tl'/N$, $\chi\rightarrow \chi/N$ in (33) does not give
all $O(1/\sqrt{N})$ terms. The $O(1)$ terms can however be found in this
way. The final expression is
\begin{eqnarray} V_{eff}&=& -{3\over 2 N\tl}\chi^2 + {1\over 2}
   \chi (\vp^2-\tv^2)+ {\Lambda\chi\over 4\pi^2}
    \nonumber \\ & &
     +{3\Lambda \tg^2\over 4\pi^2} \vp^2 -{\tg^3\over 12\pi}
     \[ (\Lambda'/2\pi^2 +\vp^2)^{3\over2}+2|\vp|^3 \]
     +{\rm const.} \end{eqnarray}
$\chi$ is given by the same solution as in the $\tg=0$ case.
Eq. (44)
is $\alpha$ independent to this order and is valid for non-negative
$\chi$, $\sqrt{\chi}$ and
$\Lambda \gg \chi, \vp^2$ and $\vp^2\gg \tg^2/(64)^2$. The matching
conditions for the linear divergences is the same as in section 3.

\noindent{\bf The case $6g^2/\l\sim O(N^{2\over3})$.}
Here we must assume $(\l  N^{2\over3})$ fixed as $N$ increases.
We can compute the leading corrections by using the replacements
$\tl=\tl'/ N^{2\over3}$, $\chi\rightarrow\chi/N^{2\over3}$ and
$\hc\rightarrow \hc/N^{2\over3}$ in the results of section 2 and
3. For the pure scalar case, $V_{Ntl}$ in eq. (11) gives at
most $O(N^{-{2\over3}})$ corrections. The rest of eq. (11) gives
$O(N^{1\over 3})$ and $O(1)$ corrections. For $g\neq 0$, the
gauge-loop ring sum included in (33) gives O(1) and also
more subdominant corrections. Thus the $O(N^{1\over 3})$+$O(1)$
corrections can then be easily found in this way. The final expression
is
\begin{eqnarray} V_{eff}&=& -{3N^{1\over3}\over 2 N^{2\over3}\tl}\chi^2 +
{N^{1\over3}\over 2}
   \chi (\vp^2-\tv^2)+
{N^{1\over3}\Lambda\chi\over 4\pi^2}-{\chi^{3\over2}
   \over 12\pi} \nonumber \\ & &
     +{3\Lambda \tg^2\over 4\pi^2} \vp^2 -{\tg^3\over 12\pi}
     \[ (\Lambda'/2\pi^2 +\vp^2)^{3\over2}+2|\vp|^3 \]
     +{\rm const.} \end{eqnarray}
$\chi$ is found from its $O(N^{1\over3})$ equation of motion, and the
comments after (44) apply. For negative $\chi$ we have to give
meaning to eq. (15). One possibility is to  simply use
$\Tr\ln(-\vd+\chi)=\Lambda\chi/2\pi^2$+const. This leads to a real
potential, but we have no rigorous way of justifying this
prescription.

\newpage
\noindent{\Large\bf 5. The Phase Transition.}

We first discuss our results generally before investigating the
phase transition.
As mentioned in the introduction, ordinary 3+1d perturbation theory
in $\hbar$ is not reliable for high temperature studies of
spontaneously broken theories because it suffers from infrared
divergence problems just near the value of the temperature at which
the phase transition from the symmetric phase to the broken symmetry
phase occurs. One can try to circumvent this problem either by
trying to resum certain infinite classes of Feynman diagrams to all
orders in $\hbar$ or by performing perturbation theory in another
parameter, say $1/N$. In order to extract reliable results
such a new perturbation series, if
it is calculable to any given order in $1/N$, must  be renormalizable
and avoid infrared divergence problems to any given order in  $1/N$.

For a weakly coupled 3+1d abelian Higgs model at high $T$ with
$6g^2/\l\leq O(1)$ we introduced an auxiliary field $\chi$ in the
Lagrangian so as to develop a systematic expansion for $V_{eff}$
in $1/N$, where $N$ is the number of real scalar fields. This
was the content of section three. The $O(N)$ and $O(1)$ corrections
in this expansion are renormalizable and avoid the infrared
divergence problems of perturbation theory in $\hbar$. The
result can be expanded to all orders in $\hbar$ and in fact
corresponds to  summing certain infinite classes of ordinary Feynman
diagrams in $\hbar$.
The O(N) result for $V_{eff}$ incorporates
the sum of all superdaisy graphs of Dolan and Jackiw [1] at O(N); the
next--to--leading order result incorporates much more, i.e. the
sum of all Feynman diagrams of O(1) as well as O(1) contributions
from the superdaisies. The $O(1)$ gauge contributions are given  by
a gauge loop ring sum (but not quite the simple one of [4]). The
beauty of our systematic approach over trying to add up some
infinite classes of diagrams by hand is that there is no overcounting
of $O(\hbar^2)$ - or any other - Feynman diagrams and in addition
the contraint of renormalizability helps to locate all relevant
contributions to $V_{eff}$ in a consistent computation.

For weak $\l$ and strong $g$ it is not possible to define an
expansion in  $1/N$, and for a weakly coupled theory it is
only possible to set $6g^2/\l\sim O(N)$ if the scalar self-coupling
is not just weak but very weak. In this case the expansion of
section 3 is unreliable because the dominant contribution is no
longer from the pure scalar fields; this is demonstrated soon.
For a very weakly  coupled 4d pure scalar field theory at high
temperature with $\l N\sim O(1)$ it is possible to write  $V_{eff}$
as the sum of $O(1)$ and $O(1/\sqrt{N})$ terms, with subdominant
terms of $O(1/N)$. In section 4 these were found to give just
the daisy sum (not superdaisy sum) of Dolan and Jackiw [1]. If
computed $consistently$ $to$ $O(1/\sqrt{N})$ the daisy sum cannot
lead to spurious field dependent terms that come from overcounting
of Feynman diagrams in $\hbar$. The daisy sum is of course
renormalizable and avoids infrared divergence problems. In the
gauged case with very weak $\lambda$ and $6g^2/\l\sim O(N)$ a
calculation of $V_{eff}$ to $O(1/\sqrt{N})$ is complicated because
the contribution from the gauge sector involves more than just the
gauge--loop ring sum of section 3. In fact, if one just adds the pure
scalar daisy sum result to $O(1/\sqrt{N})$ with the gauge--loop
ring sum to $O(1/\sqrt{N})$ then the result is multiplicatively
nonrenormalizable
in the context of the 3d field theory. We did not find a simple
expression for the $O(1/\sqrt{N})$ corrections involving gauge
fields and do not know if these corrections have infrared divergence
problems or not. The $O(1)$ corrections however have no such
problem and in addition give a gauge fixing independent effective
potential to leading order.

For $6g^2/\l\sim O(N^{2\over3})$ it is possible to easily determine
the $O(N^{1\over 3})$ and $O(1)$ contributions to $V_{eff}$, with
subdominant contributions of $O(N^{-{1\over3}})$. These were found
in section 4 and have the interpretation of the daisy sum of the
pure scalar case plus a gauge--loop ring sum taken in the limit of
vanishing (field dependent) scalar mass. The effective potential
to $O(1)$ has no infrared divergence problems, is renormalizable and
is gauge fixing parameter independent.

We now study the implications of our results for the phase transition
in the weakly coupled abelian Higgs model.
As noted in the introduction, we derived our  results by working
with an effective 3d field theory (given by eq (28)) and matching
linearly divergent terms in the effective theory with finite $O(T^2)$
corrections in the full 4d model. The ``neglected terms'' therefore
incorporate subleading terms that were dropped due to the approximations
made in calculating the 3d effective potential, as well as finite
terms that the 3d effective theory cannot account for and that arise
when the full 4d model is used. We stress that when $\chi\ll T^2$
and $\phi^2\ll T^2$, the 3d effective theory is sufficient to
calculate the leading (i.e. $T$ dependent) finite corrections.
This is of course the region of interest for the phase transition.

\noindent{\bf The case $6g^2/\l\leq O(1)$.}
Our main result is the leading + next-to-leading order high
temperature $V_{eff}(\phi^2)$ for a 3+1d abelian Higgs model
with $N$ real scalars,
computed in the systematic $1/N$ expansion. For the tree
Lagrangian given by (21) we obtained
\def\pd{\partial}
\begin{eqnarray}
 V_{eff}&=& -{3N\over 2\l}\chi^2+{N\over2}\chi(\phi^2-v^2)
     +{T^2\chi\over 24}(N+2+{18g^2\over\l})
     \nonumber \\ & &
    -{T\over12\pi}\( \(3\chi+{\l\over3}(v^2-T^2/12)\)^{3\over2}
    +(N-1)\chi^{3\over2} \) \nonumber \\
   & & -{T(6g^2/\l)^{3\over2}\over 12\pi}\[
     \(\chi+{\l\over6}(v^2+T^2/12)\)^{3\over2}  + 2
     \(\chi+{\l\over6}(v^2-T^2/12)\)^{3\over2}\] \nonumber \\
  & & + V_\alpha +\; {\rm neglected\; terms} \; + O(1/N)\;{\rm terms}.
\end{eqnarray}
where $V_{\alpha}$ is
\be V_\alpha=-{T\over12\pi}\[(\chi+\alpha g^2\phi^2)^{3\over2}
   -\chi^{3\over2}-\alpha^{3\over2}g^3 |\phi|^3 \], \ee
and $\chi(\phi)$ is the solution of the equation
$\partial V_{eff}/\partial
\chi=0$. To next--to--leading order it is sufficient [9] to use
not the full solution but the $O(N)$ solution of (19),
\be \sqrt{\chi}={\lambda T\over 48\pi}\[ \sqrt{
  1+{32\pi^2\over\l}\({12\phi^2\over T^2}-{12v^2\over T^2}+1\)}
  -1\] .\ee
There  is another solution of $\sqrt{\chi}$
which comes with an overall minus sign, but
this is unphysical [6,8,9].
Our $V_{eff}$ is valid for $T\gg\sqrt{\chi}>\l T/12\pi, g^2T/12\pi$
and $T^2\gg \phi^2-T\sqrt{\chi}/4\pi\gg g^2T^2/(64)^2$.
Also, eq. (48) was used to rewrite the $O(1)$ terms in (46), which
adds $\l (\pd V_1/\pd \chi)^2/3N$ to the $O(1/N)$ terms, where
$V_{1}$ is the $O(1)$ part of (46). In the Landau gauge, and near
$T=\sqrt{12}v$, this particular $O(1/N)$ term is negligible
for $\chi/T^2>   (\l+9g^2)(1+(6g^2/\l)^{3\over2}/2)/3 N^2$.

In our approach vacuum polarization effects
at large external momentum played an important role in
the same way as they did for the previously studied [6] $g=0$ case.

It is known [6] that at order $N$,
the potential (46) admits no 1st order phase transition. The $O(N)$
potential is exact in the small $\chi$ limit (up to 4d corrections).
To $O(N)$, $dV_{eff}/d\phi^2 = \pd V_{eff}/d\phi^2= N\chi/2$. At
$\phi^2=0$ this vanishes at $T^2_2=12v^2$. For $T>\sqrt{12}v$
the origin is a global minimum, and at $T=T_2$ the potential grows
away from the origin. For $T>T_2$ the point $\chi=0$ is away
from the origin and this has the interpretation [8,9]
as the symmetry breaking minimum below $T_2$.
For the case $g=0$, Root [9] has shown for the 3d case that the point
$\chi=0$ remains a local minimum at next--to--leading order. This
analysis did not require a computation of $V_{eff}$. Root examined
$dV/d\phi^2$ in the limit of vanishing $\chi$ and showed that the
leading order gap--equation for $\chi$ (in our case eqs. (17) and (19))
was sufficient to show that $V_{eff}/d\phi^2=0$ still has a solution
at $\chi=0$ at next--to--leading order. His analysis can be
extended to our gauged case. We will not present a detailed
analysis here but instead refer the reader to Root and
also ref. [18] where, following Root, an analysis of the
small $\chi$ limit in the full 4d high temperature abelian Higgs model
has been performed with the result that $\chi=0$ remains a
point of vanishing $dV_{eff}/d\phi^2$ at next--to--leading order.

We believe
the fact that $\chi=0$ remains a point of vanishing $dV_{eff}/d\phi^2$
is however insufficient to deduce a second order phase transition to
this order.
As mentioned by Root, for sufficiently large $N$ the $1/N$
corrections cannot overwhelm the leading order conclusion of
a second order phase transition. For $N$ not arbitrarily large
we would like to know the global properties of $V_{eff}$ away
from the point $\chi=0$, and in particular if there is
a point away from the origin at $T=\sqrt{12}v$ where the $1/N$
corrections can produce a new minimum and possibly even result
in the breakdown of the $1/N$ expansion. In addition, for $T>
\sqrt{12}v$ the point $\chi=0$ never occurs, so the analysis
of [9] is by itself insufficient to deduce $\phi^2=0$ remains
a global minimum for $N$ not arbitrarily large.  Our computation
of $V_{eff}$ sufficiently away from $\chi=0$ gives global
information that Root's analysis cannot give. In addition, (46)
appears to characterize the correct behavior in the limit of
vanishing $\chi$. Assuming no pathological behaviour occurs
in an exact computation of $V_{eff}$ at next--to--leading order
for  $\sqrt{\chi}<\lambda T/12\pi,g^2 T/12\pi$ we
might expect that (46) gives a good description all the way down
to $\sqrt{\chi}=0$ and also $\phi^2=0$ at $\sqrt{\chi}=0$.

With these points in mind
we now investigate if, more generally, (46) can exhibit a first order
phase transition for $6g^2/\l\sim O(1)$.

The critical temperature $T_2$ is given by the vanishing of
$dV/d\phi^2$ at $\phi^2=0$. Let us write $V_{eff}=V_N+V_1+V_\alpha$,
where $V_N$ is the $O(N)$ potential. Then since $\partial V_N/\partial
\chi=0$ we obtain
\begin{eqnarray}
 {dV\over d\phi^2}&=& \[{\pd V_1\over\pd\chi}+
   {\pd V_\alpha\over\pd\chi}\] {\pd\chi\over\pd\phi^2}
      +{\pd V_{eff}\over\pd\phi^2}  \nonumber \\
     &=& \[{\pd V_1\over\pd\chi}+
   {\pd V_\alpha\over\pd\chi}\]\({6\over\l}+{T\over 8\pi\sqrt{\chi}}
  \)^{-1} + {N\chi\over2}-{T\over 8\pi}
   \[\alpha g^2\sqrt{\chi+\alpha g^2\phi^2}-
      \alpha^{3\over2}g^3 |\phi|\] .\end{eqnarray}
As indicated by our preliminary remarks,
this vanishes at $\chi=0$
and at the origin this
translates to the gauge fixing independent result
\be T_2^2 = 12v^2. \ee
This is in fact the leading order result again (i.e. as if $V_1$
were absent). The reason is that in a general model the $O(1/N)$
terms in $V_{eff}$ are important in determining the $O(1/N)$ corrections
to $T_2^2$ (this is related to the fact that $\chi=0$ remains a
point of vanishing $dV_{eff}/d\phi^2$ at O(1)).

For $T>T_2$, $\phi^2=0$ is the local minimum. At $\phi^2=0$,
$V_\alpha=0$ so $V_{eff}$ is gauge parameter independent at this
minimum. For there to be a first order phase transition, eq. (49)
must have zeros for $T>T_2$ and, at $T=T_2$,
$dV/d\phi^2$ should be negative away from $\phi^2=0$. Let us work in the
Landau gauge, $\alpha\rightarrow 0$ (in an
arbitrary gauge $\alpha$ should be not more than $O(1)$ [4]).
Then at $T=T_2$,
\be {\pd V_1\over \pd\chi} =
   {T_2^2\over 12}\(1+9g^2/\l-{\sqrt{6}g\over 4\pi}(6g^2/\l) \)
-{T_2\over 8\pi}\sqrt{\chi}
    \(3\sqrt{3}-1+2(6g^2/\l)^{3\over2}\) .\ee
$\chi$ increases with $\phi^2$, hence a necessary condition for
$dV_{eff}/d\phi^2$ to become negative at $T=T_2$ is that (51) becomes
negative. At $g=0$, (51) vanishes only when $\sqrt{\chi}\sim \pi T_2
/6$ by which time the effective theory is no longer valid and the
positive contribution to $dV/d\phi^2$
from the $O(N)$ term in (49) already dominates. Hence for $g=0$
no first order phase transition is possible by $T=T_2$. For
$6g^2/\l\sim O(1)$, $g^2\ll 2\pi$,
the same conclusion can be reached, namely that within a
consistent $1/N$ approximation the leading+next-to-leading $V_{eff}$
for our weakly coupled abelian Higgs model does not admit a
first order phase transition. (Note the absence of
an  $O(|\phi|^3)$ term at $T=T_2$ -- see eq. (48).)

If $6g^2/\l$ is of $O(N)$ then (51) can become strongly negative,
indicating the possible breakdown of the $1/N$ expansion. Note
also that the ``next--to--leading'' corrections to the coefficient
of $\chi$ in (46) actually compete with the $O(N)$ coefficient
and therefore our results cannot be used in this case.

\noindent{\bf The case $6g^2/\l\sim O(N)$.} Our main result is the
high $T$ contribution to $V_{eff}$ which survives in the limit
of arbitrarily large $N$. For the tree Lagrangian given by (21)
we obtained (after eliminating $\chi$):
\begin{eqnarray}
  V_{eff} &=& {\l N\over 4!}\phi^2 (\phi^2-2v^2+T^2/6)
   + {g^2 T^2\over 8}\phi^2 \nonumber \\
   & & -{g^3 T\over 12\pi}\[ (T^2/6+\phi^2)^{3\over2}
       +2|\phi|^3 \] + O(1\sqrt{N}) + {\rm neglected}\;{\rm
      terms}.   \label{vb}
\end{eqnarray}
Note that for weak coupling one must keep $g$ and $(\l N)$ fixed
as $N$ increases. As mentioned, the potential above is $\alpha$
independent to $O(1)$ in the way we computed it. One can also
check that it is gauge fixing parameter independent with the
slightly different gauge fixing considered in [4].
Eq. (\ref{vb}) is strictly valid for $T^2\gg \phi^2\gg T^2 g^2/(64)^2$.
The critical temperature $T_2$ determined from (\ref{vb})
is given by
\be  12v^2 = T_2^2\[ 1 +3(6g^2/\l N) -{g\sqrt{3}\over\pi
    \sqrt{2}} (6g^2/\l N) \] \label{et} .\ee
We assume  $g\ll 2\pi$ so that  $T_2$ here is lower than
the leading order critical temperature for the $6g^2/\l\leq
O(N)$ case, $\sqrt{12}v$.\footnote{At the origin and at $T=T_2$ the
auxiliary field $\chi$ that was eliminated from (44) is negative, and
one may wonder if (44) and (52) are well defined in such a situation.
However, the leading divergent part of $\Tr (-\vd+\chi)^{-1}$ is
independent of $\chi$ and well defined. Hence, the linearly
divergent term in (44) is unambiguous for negative $\chi$. In fact,
to this order the quantum contribution from only the scalars
is just the $O(T^2)$
shift to the tree--level mass of the scalars already derived by
a one--loop $O(\hbar)$ calculation in [1].}
At $T=T_2$ the minimum is
no longer at the origin and occurs at the point
\be  { |\phi|_{min} \over T_2} = {1\over 2}
        g (6g^2/\l N), \label{ett} \ee
indicating a first order phase transition whose strength grows
with the gauge coupling.

Since we did not compute subleading corrections to $V_{eff}$ all
we can say is that the model exhibits a first order phase
transition for sufficiently large $N$. This is in contrast
to our results in the $6g^2/\l N\leq O(1)$ case.

\noindent{\bf The case $6g^2/\l \sim O(N^{2\over3})$. }
This case is the most interesting because, unlike the case
$6g^2/\l \sim O(N)$, the next--to--leading corrections are
easily determined. The high temperature potential is
\begin{eqnarray}
  V_{eff} &=& {\l N\over 4!}\phi^2 (\phi^2-2v^2+T^2/6)
   + {g^2 T^2\over 8}\phi^2
    - {\chi^{3\over 2}\over 12\pi} \theta(\chi)
     \nonumber \\ & &
    -{g^3 T\over 12\pi}\[ (T^2/6+\phi^2)^{3\over2}
       +2|\phi|^3 \]
     + O(N^{-{1\over 3}}) + {\rm neglected}\;{\rm
      terms}.   \label{vc}
\end{eqnarray}
Here,
\be \chi= \({\l N^{2\over3}\over 6}\)
      ( \phi^2-v^2 +T^2/12) . \ee
We note that for weak coupling, $\l N^{2\over3}$ must be held
fixed as $N$ increases and that the result for $V_{eff}$ is
gauge fixing parameter independent to $O(1)$. Finally, note the
appearence of the step function $\theta(\chi)=1$ for $\chi>0$
, 0 otherwise, in the expression for $V_{eff}$. As described in section
4 there is an uncertainty involved in defining the gaussian
integral over the quantum scalar fields when $\chi$ is negative;
we have chosen an answer that in light of the results for the
cases $6g^2/\l \leq O(1)$ and $6g^2/\l\sim O(N)$ gives a physically
appealing interpretation. Therefore, our results here are somewhat
speculative. We assume $g\ll 2\pi$ in what follows.
Also $V_{eff}$ is strictly valid for $T^2\gg \phi^2,\chi$ and
$\phi^2 \gg g^2 T^2/(64)^2 $.

The leading order critical temperature is the same as in the
case $6g^2/\l\leq O(1)$, i.e. $T_2=\sqrt{12} v$. The $O(N^{1\over3})$,
potential given by the first term in (55), does not admit a first
order phase transition.
The next--to--leading
order $T_2$ is modified. We find the solution $dV_{eff}/d\phi^2$
at $\phi^2=0$ occurs at negative $\chi$ and it is again given
by relation (\ref{et}) (but note that it reduces to $12v^2=
T_2^2$ for $N\rightarrow\infty$ here). At $T=T_2$ the global minimum
is away from the origin and at (\ref{ett}). For $6g^2/\l\sim
O(N^{2\over3})$ we have
\be  { |\phi|_{min} \over T_2} \rightarrow 0 \;\; {\rm as}\;\;
      N\rightarrow \infty, \ee
indicating a first order transition that gets weaker and weaker
as $N$ increases.

We end this section with the following discussion. If we chose a
prescription without the step function in (55) the potential is
complex at or just above $T_2$ given by (53) and close enough to
the origin. Since we have all terms to $O(1)$ the imaginary part
cannot be removed at sufficiently large $N$ by subdominant
corrections. We do not know the meaning of this result; neither
do we know  in cases a) and b) if subdominant corrections
we have not computed can lead to the possibility of a complex
potential at $T$ just above when the phase transition is supposed
to take place. However, one can check that in the case here
even without the step function in (55) the expression for $T_2^2$
is real to $O(N^{-{1\over 3}})$ and given by (53) and also
that the value of $V_{eff}$ at the minimum (54) is real to $O(1)$.
Our physical results, to the order we are working in, are
therefore not really so speculative.

\vskip 0.7cm
\noindent{\Large\bf 6. Conclusions.}

We have found that for large $N$ the weakly coupled abelian Higgs
model can exhibit either first or second order phase transitions
with first order behaviour possible for $6g^2/\l \geq O(N)$. For
$6g^2/\l\sim O(N^{2\over3})$ the model exhibits a weak first
order phase transition at large $N$. Intuitively, this only says
that the gauge coupling must be large enough so that the gauge
contributions can dominate or at least compete with the pure
scalar contributions to $V_{eff}$. For the three cases
a) $6g^2/\l\leq O(1)$, b) $6g^2/\l\sim O(N)$ and c)
$6g^2/\l \sim O(N^{2\over3})$ we estimate that our results are
reliable whenever the parameter a) $N^{-1}$, b) $N^{-{1\over2}}$
and c) $N^{-{1\over 3}}$ is sufficiently small. An inspection
of our results for case a) shows that even for $N=4$ the next
to leading corrections to the coefficient of $\phi^2$ in $V_{eff}$
compete with the leading correction to the coefficient of
$\phi^2$ in $V_{eff}$. We do not believe any of our results
can be directly applied to the $N=4$ case; however they can give
some indication of what happens at $N=4$ in a very weakly coupled
theory with $g\ll 1$, $\l \ll 1$. In this limit, cases b) and c)
admit only  a very weak first order phase transition. For case
a) it can be argued on the basis of the 3d field theory (and
dimensionality of the 3d couplings) that the coefficient of
$\phi^2$ in $V_{eff}$ to lowest order in the couplings and
to all orders in $1/N$ is given by the one--loop $O(\hbar)$
graphs. This yields a critical temperature given by [1]
\be T^2_2 (1+2/N+18g^2/\l N) = 12v^2.
 \label{ntlt} \ee
For $N=4$ and $6g^2/\l\sim 1$ this gives a critical temperature $T_2$
much below $\sqrt{12}v$ and demonstrates that our analysis of
case a) is insufficient to rule out a significant first order
phase transition. Assuming that $O(1/N)$ corrections to (46)
are negligible it is possible to use (46) to study the phase
transition to temperatures below $T=\sqrt{12}v$. The leading order
solution $\sqrt{\chi}$ of (48) becomes negative and eventually
complex near the origin as we lower the temperature below
$T=\sqrt{12}v$. To avoid this problem one can use the full solution
for $\chi$ from (46), not just the leading order one. This
approach is not entirely consistent but the result of the analysis
can be shown to be again a second order phase transition to $T
\geq T_2$ given by  (\ref{ntlt}). This only demonstrates that any
first order behaviour is contained in $O(1/N)$ corrections to
$V_{eff}$ which we can expect to be small even at $N=4$. Hence
our result for all cases is that  for a very weakly
coupled model the phase transition of
the exact model, even at $N=4$,
will be for all practical purposes second order.

For $g\sim 1$ and $N=4$, we believe that a reliable determination
of the nature of the phase transition cannot be given by the
$O(N)+O(1)$ corrections in case a), nor by the daisy and ring
sum results of cases b) and c). In all cases more subleading
contributions must be computed than those that we have computed or
currently exist in the literature. These extra contributions
may be negligible even for $g\sim 1$ and $N=4$ but this is not
$a$ $priori$ clear. However if the exact $g=0$ model admits only a
second order phase transition then we can expect that for $g\neq 0$
and $6g^2/\l\leq O(1)$ the exact model will admit at best
quite weak first order behaviour.

Our physical results are in all cases gauge fixing parameter
independent, they do not overcount any Feynman diagrams and do
not suffer from any infrared problems. We found no terms of
$O(T^3|\phi|)$ in $V_{eff}$ in any of the cases. One can argue
on the basis of the 3d field theory, dimensionality of the 3d
couplings and renormalizability that any such terms must be very
small in a very weakly coupled model. More generally, any such terms
must be vanishingly small in the limit of large $N$.

\noindent{\bf Acknowledgements.} I am very grateful to P. Weisz with
whom I had  many fruitful discussions and for a critical reading
of a draft of this paper and to
M. K. Gaillard, M. Moshe and F. Zwirner for
various useful discussions. I would also like to thank
M. Carena and C. Wagner for useful discussions and in particular
for pointing out that nonzero external momentum vacuum polarization
effects in the gauged case, as in the pure scalar case, can play
an important role. Finally, I wish to thank
K. Sibold for a clarifying discussion about
super-renormalizability. \\
\noindent{\bf Note Added.} As we completed this work we recieved
a paper by J. March--Russell, LBL--32540, PUPT--92--1328, which
also considers 2nd order phase transitions in gauge theories, but
using  $\epsilon$--expansion techniques.


\begin{thebibliography}{99}
\bibitem{pw}
As pointed out by P. Weisz, $\Tr\ln(-\vd+a\sqrt{-\vd}+b)$ can be
evaluated exactly in 3d. With a sharp momentum cutoff $\Lambda$,
and the limits $\Lambda^2 \gg b \gg a^2$ one obtains the approximate
solution $\Tr\ln(-\vd+a\sqrt{-\vd}+b)\approx \Tr\ln(-\vd+b)- ab
\ln(\Lambda^2/b)/4\pi^2$. The log divergent
term can be ignored in comparison with the linearly divergent
term whenever $(a/\Lambda)\ln(\Lambda^2/b)\ll 1$, which is clearly
true with our assumptions. The log divergent piece can be ignored in
comparison to the finite piece from $\Tr\ln(-\vd+b)$ whenever
$(a/\sqrt{b})\ln(\Lambda^2/b)\ll 1$, which is an extra assumption we
will make.
\bibitem{mss}
For $-\vd< 4\chi$, $\Pi$
is well approximated by zero. The largest error incurred by using
(39) everywhere is proportional to $\delta=\Tr\ln(-\vd+\tg^2\vp^2)-
\Tr\ln(-\vd+\tg^2\vp^2-\tg^2\sqrt{\chi}/4\pi)$ with the $\vec{k}^2$
integral from $0$ to $4\chi$. For $\vp \neq 0$, $\chi\approx 0$,
$\delta$ goes like
$\chi^{3\over2}$, i.e its contribution to the effective
mass vanishes as $\chi\rightarrow 0$. In addition, $\delta$ is
negligible for $\vp^2\gg \sqrt{\chi}/4\pi$ and $\vp^2\approx
\sqrt{\chi}/4\pi$. We will drop $\delta$ as a subdominant term,
however the fact that it goes like $\chi^{3\over2}$ for small
$\chi$ is an implicit assumtion made in the analysis in section 5.
\end{thebibliography}
\end{document}